# Indicating Asynchronous Array Multipliers

P. Balasubramanian, D.L. Maskell

*Abstract*—Multiplication is an important arithmetic operation that is frequently encountered in microprocessing and digital signal processing applications, and multiplication is physically realized using a multiplier. This paper discusses the physical implementation of many indicating asynchronous array multipliers, which are inherently elastic and modular and are robust to timing, process and parametric variations. We consider the physical realization of many indicating asynchronous array multipliers using a 32/28nm CMOS technology. The weak-indication array multipliers comprise strong-indication or weak-indication full adders, and strong-indication 2-input AND functions to realize the partial products. The multipliers were synthesized in a semi-custom ASIC design style using standard library cells including a custom-designed 2-input C-element. 4×4 and 8×8 multiplication operations were considered for the physical implementations. The 4-phase return-to-zero (RTZ) and the 4-phase return-to-one (RTO) handshake protocols were utilized for data communication, and the delay-insensitive dual-rail code was used for data encoding. Among several weak-indication array multipliers, a weak-indication array multiplier utilizing a biased weak-indication full adder and the strong-indication 2-input AND function is found to have reduced cycle time and power-cycle time product with respect to RTZ and RTO handshaking for 4×4 and 8×8 multiplications. Further, the 4-phase RTO handshaking is found to be preferable to the 4-phase RTZ handshaking for achieving enhanced optimizations of the design metrics.

## I. Introduction

MULTIPLICATION is an important arithmetic operation that is frequently encountered in microprocessing and digital signal processing [1], [2]. References [3–9] discuss various transistor-level and gate-level designs of the asynchronous multipliers. However, a majority of these multipliers correspond to the bundled-data handshake protocol, which has separate request and acknowledge wires besides the data bundle (i.e., data bus) and features a constant delay element that governs data communication between the sender and the receiver. Due to the fixed delay presumed for the data transfer between the sender and the receiver, bundled-data asynchronous multipliers are not robust when the presumed delay gets exceeded, and they are neither indicating nor robust.

This work is supported by the Academic Research Fund Tier-2 research award of the Ministry of Education, Republic of Singapore under Grant MOE2017-T2-1-002.

P. Balasubramanian and D.L. Maskell are with the School of Computer Science and Engineering, Nanyang Technological University, Singapore 639798 (e-mails: balasubramanian@ntu.edu.sg, asdouglas@ntu.edu.sg).

In this work, we consider the robust class of indicating asynchronous multipliers whose product bits acknowledge the arrival of all the primary inputs and the completion of internal computation. Indicating asynchronous circuits are quasi-delay-insensitive circuits, which are the practically realizable delay-insensitive circuits which include the weakest compromise of isochronic fork(s) [10]. All the wires branching out from an isochronic node or junction are assumed to experience signal transitions i.e., rising or falling concurrently. In this work, we consider the array multiplier architecture for an example, which corresponds to the well-known shift-and-add multiplication approach. We realize indicating asynchronous realizations of 4×4 and 8×8 array multipliers, which utilize asynchronous components pertaining to strong-indication and weak-indication asynchronous logic design methods.

The rest of the article is organized into 4 sections. Section 2 gives background information about the design of indicating asynchronous circuits. Section 3 discusses various indicating asynchronous implementations of the 4×4 and 8×8 array multipliers by following the semi-custom ASIC design style. Section 4 presents the design metrics estimated for the array multipliers based on physical realization using a 32/28nm CMOS process. The (normalized) power-cycle time product of the multipliers is also provided. Finally, some conclusions and a scope for further work are mentioned in Section 5.

## II. Indicating Asynchronous Circuits – A Background

### A. Data Encoding and Handshaking

The schematic of an indicating asynchronous circuit stage is shown in Fig. 1, which is correlated with the sender-receiver analogy. In Fig. 1, the current stage and the next stage registers are analogous to the sender and the receiver, and the indicating asynchronous circuit is sandwiched between the current stage and the next stage register banks. The register bank comprises a series of registers, with one register allotted for each of the rails of an encoded data input. Here, the register is basically a 2-input C-element. The C-element will output 1 or 0 if all its inputs are 1 or 0 respectively. If the inputs to a C-element are not identical then the C-element would retain its existing steady-state. The circles with the marking 'C' denote the C-elements in the figures.

In Fig. 1, (Q1, Q0), (R1, R0) and (S1, S0) represent the delay-insensitive dual-rail encoded inputs of the single-rail inputs Q, R and S respectively. According to dual-rail data encoding [11] and 4-phase RTZ handshaking [12], an input Q is encoded as (Q1, Q0) where Q = 1 is represented by Q1 = 1 and Q0 = 0, and Q = 0 is represented by Q0 = 1 and Q1 = 0.

Both these assignments are called data. The assignment Q1 = Q0 = 0 is called the spacer, and the assignment Q1 = Q0 = 1 is deemed to be illegal since the coding scheme should be unordered [13] to maintain the delay-insensitivity.

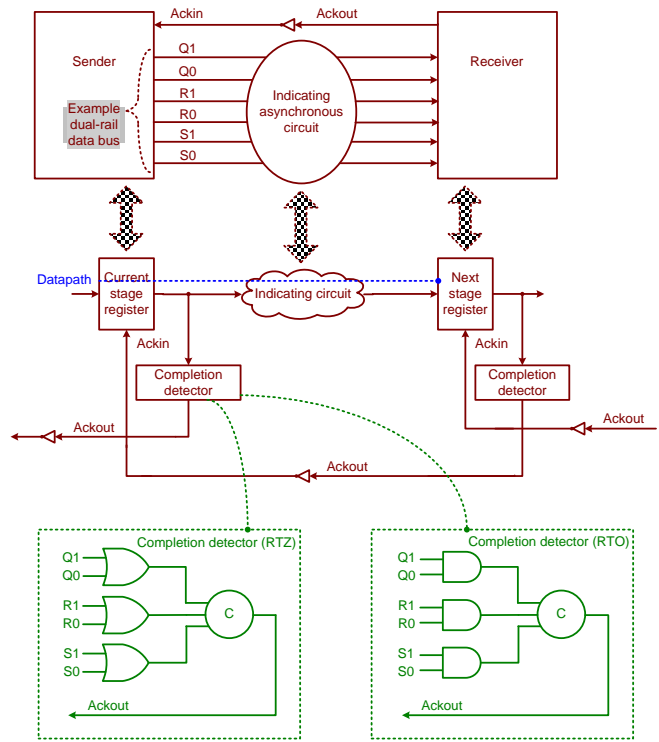

Fig. 1 A indicating asynchronous circuit stage. Example RTZ and RTO completion detectors are portrayed within the dotted green boxes in green.

The application of input data to a indicating asynchronous circuit obeying the 4-phase RTZ handshaking follows this sequence [12]: data-spacer-data-spacer, and so forth. The application of input data is followed by the application of the spacer, which implies that there is an interim RTZ phase between the successive applications of input data. The RTZ phase ensures an unambiguous data communication between the sender and the receiver. The RTZ handshaking is governed by the following four steps.

- Step 1: The dual-rail data bus specified by (Q1, Q0), (R1, R0) and (S1, S0) assumes the spacer, and therefore the acknowledgment input (Ackin) is binary 1. After the sender transmits a data, this would cause rising signal transitions i.e., binary 0 to 1 to occur on one of the dual rails of the dual-rail data bus
- Step 2: The receiver would receive the data sent and drive the acknowledgment output (Ackout) to 1
- Step 3: The sender waits for Ackin to become 0 and would reset the dual-rail data bus, i.e., the dual-rail data bus becomes the spacer again
- Step 4: After an unbounded but a finite and positive time duration, the receiver drives Ackout to 0 and subsequently Ackin would assume 1. With this, a single data transaction is said to be completed, and the asynchronous circuit is permitted to start the next data transaction

According to dual-rail data encoding and 4-phase RTO handshaking [14], an input Q is encoded as (Q1, Q0) and Q = 1 is represented by Q1 = 0 and Q0 = 1, and Q = 0 is represented by Q0 = 0 and Q1 = 1. Both these assignments are called data. The assignment Q1 = Q0 = 1 is called the spacer, and the assignment Q1 = Q0 = 0 is deemed to be illegal to maintain the delay-insensitivity.

The application of input data to a indicating asynchronous circuit obeying the 4-phase RTO handshaking follows this sequence: spacer-data-spacer-data, and so forth. There is an interim RTO phase between the successive applications of input data and the RTO phase ensures an unambiguous data communication between the sender and the receiver. The RTO handshaking process is governed by the following four steps.

- Step 1: Ackin is equal to binary 1. After the sender transmits the spacer, this would cause rising signal transitions i.e., binary 0 to 1 on all the rails of the dual-rail data bus
- Step 2: The receiver would receive the spacer sent and drive Ackout to 1
- Step 3: The sender waits for Ackin to become 0 and would transmit the data through the dual-rail data bus
- Step 4: After an unbounded but a finite and positive time duration, the receiver drives Ackout to 0 and subsequently Ackin would assume 1. With this, a single data transaction is said to be completed, and the asynchronous circuit is permitted to start the next data transaction

In a indicating asynchronous circuit, the time taken to process the data in the datapath, highlighted by the dotted blue in Fig. 1, is called forward latency and the time taken to process the spacer is called reverse latency. Since there is an intermediate RTZ or RTO phase between the application of two input data sequences, the cycle time is expressed by the sum of forward and reverse latencies. The cycle time of a indicating asynchronous circuit is synonymous with the clock period of a synchronous circuit.

The gate-level detail of the example completion detectors corresponding to 4-phase RTZ and RTO handshake protocols are shown at the bottom of Fig. 1, within the dotted green boxes. A completion detector acknowledges the receipt of all the primary inputs given to an indicating asynchronous circuit stage. In the case of 4-phase RTZ handshaking, Ackout is provided by employing a 2-input OR gate to combine the respective dual rails of each encoded input, and then synchronizing the outputs of such 2-input OR gates using a C-element or a tree of C-elements. In the case of 4-phase RTO handshaking, Ackout is provided by employing a 2-input AND gate to combine the respective dual rails of each encoded input, and then synchronizing the outputs of such 2-input AND gates using a C-element or a tree of C-elements. It may be noted that Ackin is the Boolean complement of Ackout and vice-versa.

*B. Indicating Asynchronous Circuits*

Indicating asynchronous circuits are classified into strong-indication and weak-indication circuits [15]. The input-output timing correlation of these circuit types is illustrated by a representative timing diagram shown in Fig. 2. The arrival of data or spacer is highlighted within the dotted blue circles and the complete receipt of data or spacer is highlighted within the dotted green ovals in Fig. 2.

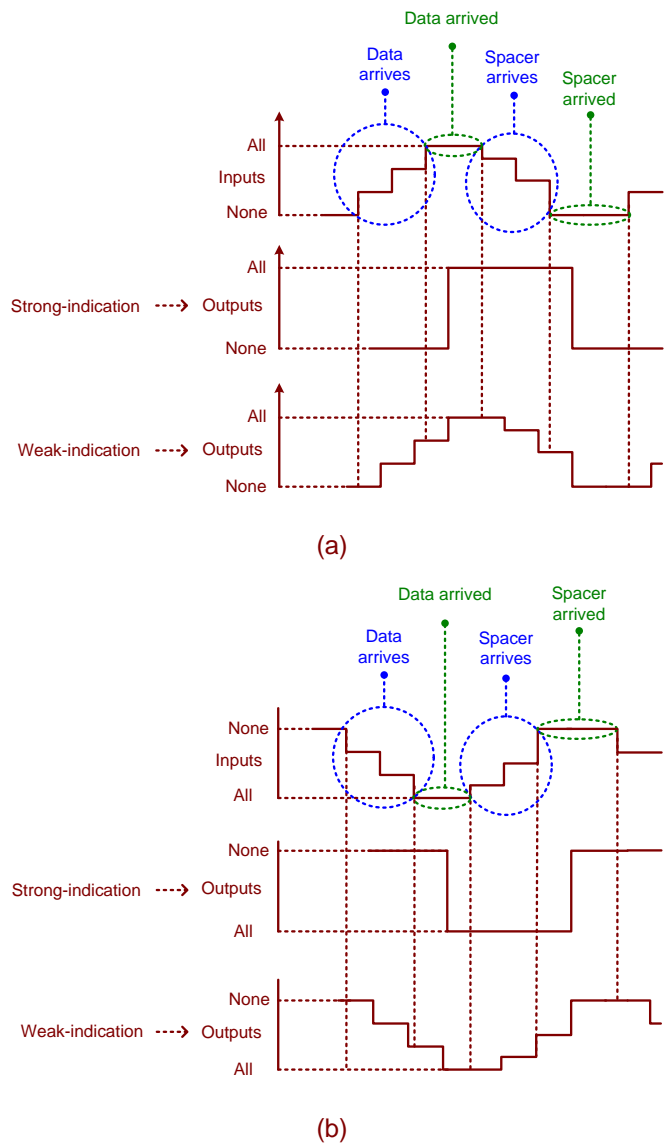

(a)

(b)

Fig. 2 Input-output timing correlation of strong-indication and weak-indication circuit types corresponding to: (a) RTZ handshaking and (b) RTO handshaking.

Strong-indication circuits [16] would wait to receive all the primary inputs (i.e., data or spacer) and would then process them to produce the required primary outputs (data or spacer).

On the other hand, weak-indication circuits [17] can produce all but one of the primary outputs after receiving a subset of the primary inputs. Nevertheless, only after receiving the last primary input, they would produce the last primary output.

Both strong- and weak-indication asynchronous circuit types incorporate the isochronic fork assumption [10]. It is reported in [18] that the isochronic fork assumption is also realizable in the nanoelectronics regime.

A cascade of strong-indication sub-circuits may not result in a strong-indication circuit. Rather, a weak-indication circuit may result. For example, if two strong-indication full adders are cascaded, the resultant would be a weak-indication 2-bit ripple carry adder (RCA). This is because if all the inputs to one of the full adders are provided, the corresponding sum and carry output bits of that full adder could be produced regardless of the provision of inputs for the other full adder in the RCA. However, only after the inputs to the other full adder are supplied, its corresponding sum and carry output bits would be produced. This is characteristic of weak-indication.

Among the strong- and weak-indication circuits, the latter are preferable for physical implementation [19]. This is because of the strict timing restrictions inherent in the former. Generally, for implementing arithmetic functions, the weak-indication type is preferable to the strong-indication type [20–22] and this is because strong-indication arithmetic circuits tend to experience worst-case forward and reverse latencies for the application of data and spacer, and therefore the cycle time of strong-indication arithmetic circuits is always the maximum.

On the other hand, weak-indication arithmetic circuits may encounter data-dependent forward and reverse latencies or a data-dependent forward latency and a constant reverse latency. Thus, the cycle time of weak-indication arithmetic circuits is usually less compared to that of strong-indication arithmetic circuits. However, for the weak-indication array multipliers considered here it is observed that their forward and reverse latencies are neither data-dependent nor a constant; rather they correspond to the worst-case timing and so the cycle time also corresponds to the worst-case. Nevertheless, it is noted that the weak-indication array multipliers incorporating weak-indication full adders facilitate reductions in cycle time, silicon area, and average (total) power dissipation compared to the weak-indication array multipliers incorporating strong-indication full adders. This will be evident from the simulation results presented in Section 4.

III. INDICATING ASYNCHRONOUS ARRAY MULTIPLIERS

Many weakly indicating 4×4 and 8×8 array multipliers were physically implemented corresponding to RTZ and RTO handshaking. References [23], [24], [41] provide practical examples for the transformation of an asynchronous circuit corresponding to the RTZ protocol into that that corresponds to the RTO protocol and vice-versa. The rules for the logical transformation between RTZ and RTO handshaking are given in [25] along with the proofs, and an interested reader may refer to the same for details. However, the example RTZ and RTO completion detectors shown in Fig. 1 serve as a small illustration for the logic transformation between RTZ and RTO handshaking. Another example of such a logic transformation is portrayed by Fig. 3, which shows the 2-input AND function realized according to RTZ and RTO handshaking.

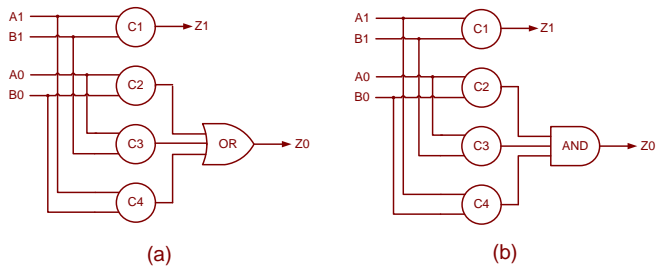

Fig. 3 Strongly indicating realization of 2-input AND function corresponding to: (a) RTZ handshaking and (b) RTO handshaking. C1 to C4 represent the 2-input C-elements in (a) and (b). (A1, A0) and (B1, B0) are the inputs and (Z1, Z0) is the output.

Note that robust asynchronous realizations of the 2-input AND function are required to generate the partial products. For this purpose, a strongly indicating realization of the 2-input AND function is considered as shown in Fig. 3. It may be noted in this context that a weak-indication 2-input AND function cannot be realized since it has only one dual-rail primary output. A weak-indication circuit requires at least a pair of dual-rail primary outputs to satisfy the weak-indication timing constraints.

The primary intent here is to determine which indicating asynchronous logic components would be more optimum for realizing the array multiplier architecture. This observation may be useful to determine which indicating asynchronous logic components would be more suitable for the optimum realization of indicating asynchronous multipliers based on the other multiplier architectures. Further, it is of interest to ascertain whether the RTZ or the RTO handshaking could help to better optimize the design metrics.

The indicating full adders derived based on different asynchronous logic design methods [26–31] are used to realize the asynchronous array multipliers, as mentioned below, by substituting the full adders in the array multiplier architectures shown in Figs. 4a and 4b. Strongly indicating realizations of the 2-input AND function, as shown in Fig. 3, were used to generate the partial products to perform shifted-addition.

- Weak-indication array multipliers which incorporate strong-indication full adders realized based on [26] corresponding to RTZ and RTO handshaking
- Weak-indication array multipliers which incorporate strong-indication full adders realized based on [27] corresponding to RTZ and RTO handshaking
- Weak-indication array multipliers which incorporate strong-indication full adders realized based on [28] corresponding to RTZ and RTO handshaking
- Weak-indication array multipliers which incorporate weak-indication full adders realized based on [27] corresponding to RTZ and RTO handshaking
- Weak-indication array multipliers which incorporate weak-indication full adders realized based on [29] corresponding to RTZ and RTO handshaking
- Weak-indication array multipliers which incorporate weak-indication full adders realized based on [30] corresponding to RTZ and RTO handshaking
- Weak-indication array multipliers which incorporate weak-indication full adders realized based on [31] corresponding to RTZ and RTO handshaking

IV. SIMULATION RESULTS

Twenty-eight indicating asynchronous array multipliers corresponding to 4×4 and 8×8 multiplications were physically realized using the gates of a 32/28nm CMOS standard digital cell library [32]. In our previous work [42], only the 4×4 multiplication was considered. The 2-input C-element does not form a part of the cell library and so it was custom-realized based on the AO222 cell by introducing feedback, which required 12 transistors. All the asynchronous array multipliers correspond to weak-indication. An N×N array multiplier gives rise to $N^2$ partial products, which are realized using 2-input AND functions, and then shifted and added using N(N–1) full adders with the carry input of N full adders reset (i.e., set to 0 and 1 in the case of RTZ and RTO handshaking respectively).

Quasi-delay-insensitivity was carefully considered while decomposing the asynchronous logic [28] [33] to avoid the possibility of creation of gate orphan(s). Gate orphans are unacknowledged signal transitions on the intermediate gate outputs, which are problematic as they might affect the robustness of an indicating asynchronous circuit and so they should be avoided [34]. For an explanation of gate orphans, we refer the readers to some prior works [35–37]. Wire orphan refers to the unacknowledged signal transition on a wire and is avoided by imposing the isochronic fork assumption [38].

A typical case PVT specification of the high $V_t$ digital cell library [32] with a supply voltage of 1.05V and an operating junction temperature of 25°C was considered to perform the simulations. The design metrics such as cycle time, area, and average power dissipation estimated are given in Table I.

The cycle time of an indicating asynchronous circuit is the sum of forward and reverse latencies. The forward latency is like the critical path delay which can be directly estimated. The estimation of reverse latency is non-trivial since it represents the time taken to process the spacer and the reverse latency cannot be directly estimated using a commercial static timing analyzer. The reverse latency can however be estimated based on the timing information obtained through the gate-level simulations. For the indicating asynchronous multipliers in Table I, their forward and reverse latencies are equal, and so the cycle time is easily estimated. This is because the longest datapath traversed for the application of data or spacer is the same as highlighted by the dotted blue lines in Figs. 4a and 4b.

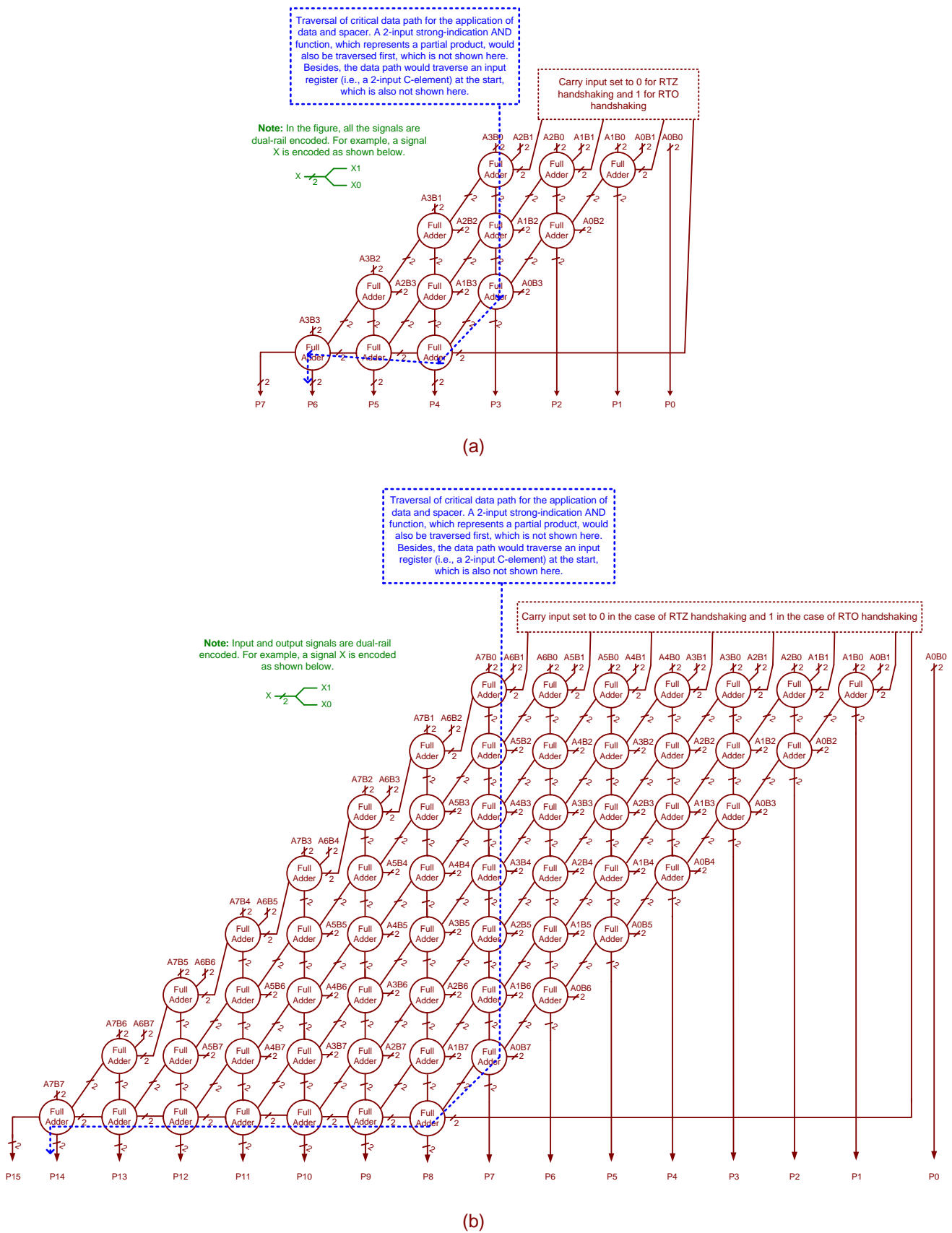

Fig. 4 Schematics of: (a) 4×4 array multiplier and (b) 8×8 array multiplier. The primary inputs, partial products, intermediate outputs, and primary outputs are all dual-rail encoded. The critical paths are highlighted by the dotted blue lines in (a) and (b). P7 to P0 represent the product bits of the 4×4 array multiplier, and P15 to P0 represent the product bits of the 8×8 array multiplier.

Since power and cycle time are desired to be less, the power-cycle time product (PCTP) should also be less. The PCTP serves as a qualitative low power/energy metric for an indicating asynchronous circuit design, which is analogous to the power-delay product of a synchronous circuit design.

The PCTP of the indicating asynchronous multipliers were calculated and normalized. To perform the normalization, the highest value of PCTP of an array multiplier corresponding to a particular multiplication operation (i.e., 4×4 or 8×8) was considered as the reference, and this was used to divide the actual PCTPs of the other array multipliers corresponding to the same multiplication operation. This procedure was adopted to obtain the normalized PCTP values of the asynchronous array multipliers corresponding to RTZ and RTO handshaking. Hence, the least value of PCTP corresponding to an indicating asynchronous array multiplier for a specific multiplication is representative of the best design with respect to RTZ/RTO handshaking.

TABLE I.   DESIGN PARAMETERS OF 4×4 AND 8×8 INDICATING ASYNCHRONOUS MULTIPLIERS, ESTIMATED USING A 32/28-NM CMOS PROCESS

| Multiply Operation | Literature Reference | Cycle Time (ns) | Area (µm²) | Power (µW) | PCTP (Norm.) |
|---|---|---|---|---|---|
| *Corresponding to RTZ handshaking* | | | | | |
| 4×4 | [26] | 7.26 | 1015.30 | 1245 | 1 |
| | [27][1] | 5.42 | 1006.16 | 1228 | 0.736 |
| | [28] | 5.32 | 926.86 | 1207 | 0.710 |
| | [27][2] | 5.20 | 975.66 | 1222 | 0.703 |
| | [29] | 5.18 | 823.17 | 1216 | 0.697 |
| | [30] | 3.90 | 853.67 | 1222 | 0.527 |
| | [31] | 4.48 | 835.37 | 1217 | 0.603 |
| *Corresponding to RTO handshaking* | | | | | |
| 4×4 | [26] | 7.08 | 1015.31 | 1240 | 1 |
| | [27][1] | 5.16 | 957.36 | 1211 | 0.712 |
| | [28] | 5.24 | 926.86 | 1206 | 0.720 |
| | [27][2] | 5.02 | 951.26 | 1211 | 0.692 |
| | [29] | 5.12 | 823.17 | 1212 | 0.707 |
| | [30] | 3.70 | 853.67 | 1217 | 0.513 |
| | [31] | 4.38 | 835.37 | 1213 | 0.605 |
| *Corresponding to RTZ handshaking* | | | | | |
| 8×8 | [26] | 14.52 | 4259.19 | 1532 | 1 |
| | [27][1] | 11.08 | 4216.50 | 1491 | 0.743 |
| | [28] | 11.92 | 3846.46 | 1461 | 0.783 |
| | [27][2] | 10.56 | 4074.18 | 1477 | 0.701 |
| | [29] | 10.32 | 3362.58 | 1461 | 0.678 |
| | [30] | 8.40 | 3504.90 | 1474 | 0.557 |
| | [31] | 8.66 | 3419.50 | 1461 | 0.569 |
| *Corresponding to RTO handshaking* | | | | | |
| 8×8 | [26] | 14.14 | 4259.19 | 1514 | 1 |
| | [27][1] | 10.54 | 3988.79 | 1444 | 0.711 |
| | [28] | 11.78 | 3846.46 | 1452 | 0.799 |
| | [27][2] | 10.22 | 3960.32 | 1443 | 0.689 |
| | [29] | 10.20 | 3362.58 | 1447 | 0.689 |
| | [30] | 8 | 3504.90 | 1459 | 0.545 |
| | [31] | 8.50 | 3419.50 | 1449 | 0.575 |

[1] Uses strong-indication full adder; [2] Uses weak-indication full adder

It can be seen from Table I that the average power dissipation does not vary significantly across the array multipliers corresponding to a particular handshaking, and this is because all the indicating asynchronous array multipliers satisfy the monotonic cover constraint (MCC) [11]. The MCC basically refers to the activation of a unique signal path from a primary input to a primary output for the application of an input data. The MCC arises from the adoption of a logic expression format which is composed of disjoint or orthogonal terms [39] to describe the primary outputs. For example, in a disjoint sum-of-products expression, the logical conjunction of any two product terms would yield zero [40], [43]. Hence, only one term gets activated in a disjoint logic expression subsequent to the application of an input data. Incorporating the MCC ensures the proper indication of signal transitions throughout an asynchronous circuit over the entire circuit depth from the first up to the last logic level. This is because the signal transitions, whether they be rising or falling, should occur monotonically throughout an indicating asynchronous circuit [34], and satisfying the MCC and performing quasi-delay-insensitive logic decomposition guarantees this.

Two important observations can be made from Table I. Firstly, the weak-indication array multiplier incorporating the biased weak-indication full adder of [30] and the strong-indication 2-input AND function (to realize the partial products) reports less cycle time and PCTP compared to its counterparts with respect to RTZ and RTO handshaking for both 4×4 and 8×8 multiplications. Amongst the counterpart designs, the weak-indication array multiplier featuring the weak-indication full adder of [31] and the strong-indication 2-input AND functions for the partial products is better than the rest with respect to 4×4 and 8×8 multiplications. Secondly, the RTO handshaking enables consistent reductions in the design metrics compared to the RTZ handshaking.

With respect to 4×4 multiplication, compared to the weak-indication array multiplier constructed using the weak-indication full adder of [31], the weak-indication array multiplier embedding the weak-indication full adder of [30] reports reductions in PCTP by 12.6% for RTZ handshaking and 15.2% for RTO handshaking. With respect to 8×8 multiplication, the weak-indication array multiplier incorporating the weak-indication full adder of [30] reports respective reductions in PCTP by 2.1% and 5.2% for RTZ and RTO handshaking compared to the weak-indication array multiplier incorporating the weak-indication full adder of [31].

As the size of the multiplication is increased from 4×4 to 8×8, it is noticed that the weak-indication array multiplier based on [30] achieves less reductions in cycle time and PCTP compared to the weak-indication array multiplier based on [31]. This is mainly due to the increased datapath delay experienced in the critical path of the former compared to the latter. For example, considering RTZ handshaking, the critical path through the full adder of [30] which corresponds to the carry output involves an AO222 gate, and the critical path through the full adder of [31] which corresponds to the carry output involves one AO21 gate. The typical propagation delay of the (minimum-sized) AO21 gate being 41% less than the typical propagation delay of the (minimum-sized) AO222 gate [32]. Hence, it may be that as the size of the multiplication is further increased from 8×8, the weak-indication array multiplier based on [31] might become competitive to the weak-indication array multiplier based on [30].

## V. Conclusions and Further Work

This article has discussed the physical implementation of robust indicating asynchronous array multipliers, based on the 4-phase RTZ and RTO handshake protocols. The asynchronous array multipliers are robust and correspond to the weak-indication timing model. The consideration of the array multiplier is motivated by the fact that it has a very regular layout and hence it is easy to pipeline if any increase in the throughput is desired.

It is inferred that the weak-indication array multiplier incorporating the weak-indication full adder of [30] enables enhanced optimizations in the design metrics compared to the other weak-indication array multipliers utilizing other full adders such as [26–29], [31]. As the size of the multiplication is further increased, we hypothesize that the array multiplier utilizing the weak-indication full adder of [31] might become competitive to that utilizing the weak-indication full adder of [30]. However, a practical implementation and analysis are necessary. Nevertheless, both [30] and [31] present the designs of full adders which incorporate redundant logic, and it was shown in [44] that redundant logic could help to significantly reduce the latencies and the cycle time and associated with just negligible increases in area and average power dissipation.

Construction of indicating asynchronous array multipliers as given in Table I is quite straightforward since the full adders based on the corresponding design methods [26–31] can be placed directly in the architectures shown in Figs. 4a and 4b, corresponding to RTZ or RTO handshaking. However, the construction of robust early output asynchronous array multipliers using the early output full adders of [22], [45] and [46] may not be straightforward. This is due to the likelihood of the problem of gate orphans. To overcome the gate orphan problem in the realization of an early output asynchronous array multiplier, the provision of internal completion detectors, as discussed in [47], may become necessary to ensure full indication of the rising and falling signal transitions at the intermediate gate outputs. Moreover, the outputs of all the internal completion detectors have to be synchronized with at least one dual-rail product bit of the array multiplier using a tree of C-elements. This would enable the provision of proper acknowledgment for the receipt of data or spacer starting from the first logic level up to the last logic level.

Although the early output logic better optimizes the physical realization of the full adders thereby suggesting potential savings in the design metrics of asynchronous array multipliers the additional introduction of internal completion detectors for quasi-delay-insensitivity might offset the reductions in the design metrics achieved based on the early output logic. This could be a subject matter for future investigation. Hence, the design and implementation of robust early output asynchronous array multipliers and their comparative evaluation with indicating asynchronous multipliers in terms of the design metrics is necessary, which suggests a scope for further work.